\documentclass[conference]{IEEEtran}
\IEEEoverridecommandlockouts
\usepackage{amsmath,amsfonts}
\usepackage{xcolor}
\usepackage{amsthm}
\usepackage{algorithmic}
\usepackage{algorithm}
\usepackage{array}
\usepackage[caption=false,font=normalsize,labelfont=sf,textfont=sf]{subfig}
\usepackage{textcomp}
\usepackage{stfloats}
\usepackage{hyperref}
\usepackage{balance}
\usepackage{url}
\usepackage{verbatim}
\usepackage{graphicx}
\usepackage{epstopdf}
\usepackage{cite}
\usepackage{booktabs}    
\usepackage{amssymb}     
\usepackage{graphicx}    

\begin{document}
\title{A Deep Q-Network based power control mechanism to Minimize RLF driven Handover Failure in 5G Network}
\author{\IEEEauthorblockN{Kotha Kartheek$^{a}$, Shankar K. Ghosh$^{a}$, Megha Iyengar$^{a}$, Vinod Sharma$^{b}$} 
\IEEEauthorblockA{$^{a}${\it Department of Computer Science and Engineering} \\
$^{b}${\it Department of Electrical Engineering} \\
Shiv Nadar Institution of Eminence \\
Delhi NCR, India\\
Emails: \{kk746, shankar.ghosh, mk197, vinod.sharma\}@snu.edu.in} 
\and
\IEEEauthorblockN{Souvik Deb}
\IEEEauthorblockA{{\it ACM Unit} \\
\textit{Indian Statistical Institute}\\
Kolkata, India \\
Email: deb.souvik5@gmail.com}
}

\maketitle

\begin{abstract}
The impact of Radio link failure (RLF) has been largely ignored in designing handover algorithms, although RLF is a major contributor towards causing handover failure (HF). RLF can cause HF if it is detected during an ongoing handover. The objective of this work is to propose an efficient power control mechanism based on Deep Q-Network (DQN), considering handover parameters (i.e., time-to-preparation, time-to-execute, preparation offset, execution offset) and radio link monitoring parameters (T310 and N310) as input. The proposed DRL based power control algorithm decides on a possible increase of transmitting power to avoid RLF driven HF. Simulation results show that the traditional conditional handover, when equipped with the proposed DRL based power control algorithm can significantly reduce both RLFs and subsequent HFs, as compared to the existing state of the art approaches.   
\end{abstract}


\IEEEpeerreviewmaketitle

\section{Introduction}

To sustain connectivity with a New Radio (NR) system, user equipments (UEs) have to switch from one Next Generation Node B (gNB) to another. This is known as \emph{handover} \cite{refhandover}. Typically, handover decision in NR is made based on some parameters such as time-to-execute ($T_{exec}$), time-to-preparation ($T_{prep}$), preparation offset ($O_{prep}$) and execution offset ($O_{exec}$) \cite{CHODeb}. In the widely known conditional handover for NR systems, the handover process is initiated upon meeting the following condition for handover preparation \cite{CHODeb}:  
\begin{equation} \label{condition1}
    P_t  > P_c + O_{prep}, \; \text{for $T_{prep}$ period of time.}
\end{equation}
\begin{figure}[htb]
    \centering
    \includegraphics[scale=0.45]{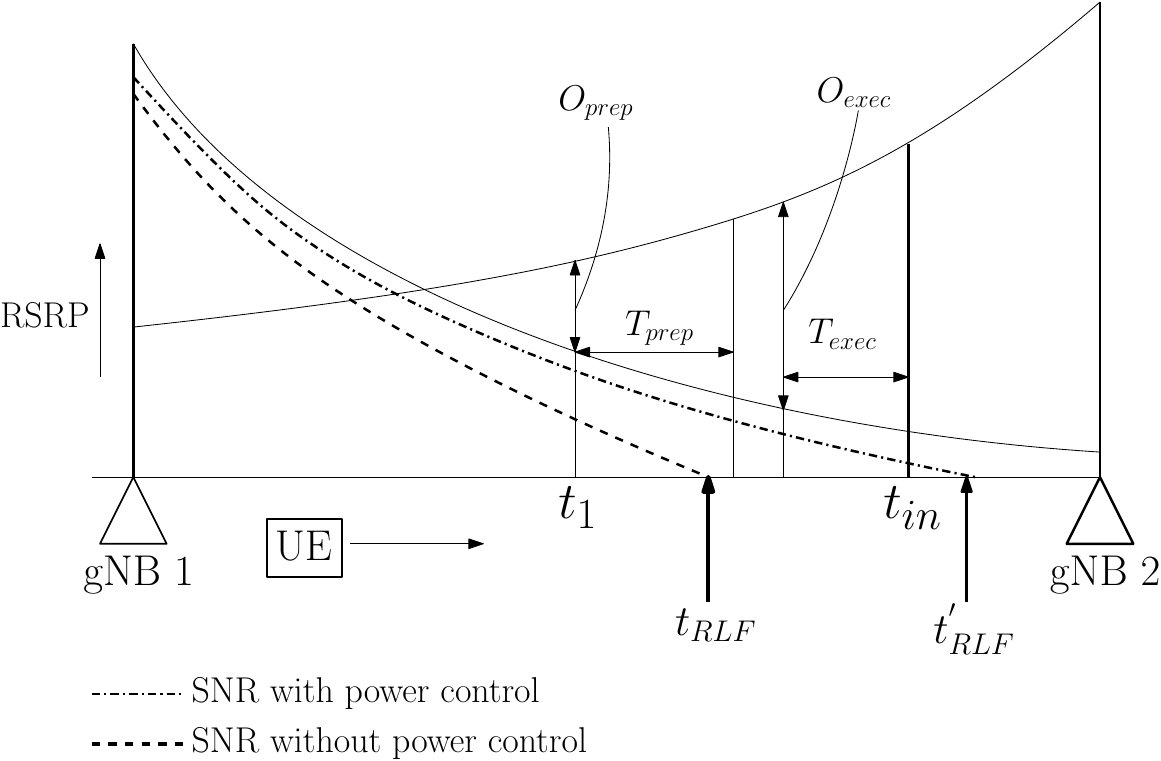}
    \caption{2-gNB model demonstrating RLF induced HF.}
    \label{model}
\end{figure}
\noindent i.e., $P_t$, the downlink reference signal received power (RSRP) from the neighboring gNB is greater than the downlink RSRP from the serving gNB by $O_{prep}$ amount for all RSRP sampling instances (taken in every 200 ms \cite{lopez}) during $T_{prep}$. After handover preparation phase, the handover execution phase starts. The handover execution phase is successful upon meeting the following condition \cite{CHODeb} (depicted in Fig. \ref{model}):
\begin{equation} \label{condition2}
    P_t  > P_c + O_{oxec}, \; \text{for $T_{exec}$ period of time}.
\end{equation}
An inappropriate setting of handover parameters, i.e., $T_{prep}$, $T_{exec}$, $O_{prep}$ and $O_{exec}$, will make the UE to wait longer before the handover is executed. In the mean time, the signal to noise ratio (SNR) from the current gNB may degrade severely resulting in radio link failure (RLF) \cite{lopez}. It may be noted that the occurrence of RLF is regulated by two parameters namely T310 and N310. An UE is considered to be out of synchronization (out-of-sync) when its SNR falls below a predefined threshold ($S_{RLF}$). The T310 timer is triggered if the UE encounters N310 consecutive out-of-sync events. The UE is back to synchronization if the SNR increases above the in-sync threshold ($Q_{in}$), and the T310 timer stops. However, if the T310 timer runs until expiration, the UE is considered to be out of synchronization, and an RLF is declared \cite{lopez}.
As per the definition of 3GPP, if RLF is detected when the Time-to-trigger (TTT) timer is running, \emph{handover failure} (HF) is declared by the gNB \cite{3gpp}. High HF results in higher handover latency which is quiet unacceptable for delay stringent services \cite{NRURllC}. In order to minimize RLF driven HF, the RLF event need to be avoided during an ongoing handover. To illustrate, let us consider the 2-gNB model depicted in Fig. \ref{model}.  We consider that an UE (connected to gNB 1) is moving from gNB 1 to gNB 2 through a linear trajectory. Here handover failure is demonstrated for two different transmitting power levels of gNB 1. As the UE is moving from gNB 1 to gNB 2, the $T_{prep}$ timer is started at time $t_1$; and the handover decision is made as soon as the $T_{exec}$ timer expires (at time $t_{in}$). In the mean time, SNR from gNB 1 degrades, and RLF is declared at $t_{RLF}$, i.e., $t_1 \leq t_{RLF} \leq t_{in}$. Such an RLF results in HF. On the other hand, for an increased transmitting power, the RLF event is deferred till $t_{RLF}^{'}$ and handover is executed beforehand (at $t_{in}$). As a consequence, HF is avoided. Additionally, HF depends on UE velocity as well. For example, if the UE velocity is very high, then the UE will move away from gNB 1 rapidly during $t_{in}-t_{1}$, resulting in RLF and subsequent HF. Moreover the absolute distance of the UE determining the RSRP level depends on the trajectory of the UE, and thereby playing a crucial role in causing RLF and subsequent HF. {\bf In summary, an efficient power control mechanism considering the handover parameters ($T_{exec}$, $T_{prep}$, $O_{exec}$ and $O_{prep}$), RLF parameters (N310, T310), UE velocity, distances of the UE from the current and target gNBs, RSRP levels at the UE from current and target gNBs are required to minimize RLF and subsequent HF.}

The Markov model analysis of HF in \cite{CHODeb} considering handover parameters do not account for the transmit power control at the gNB to minimize RLF driven HF. The logistic regression method in \cite{refhandover} to predict the possible occurrence of a handover considers RSRP from all gNBs, recived signal strength indicator (RSSI) at the UE, hysteresis, TTT  and distance of the UE from the serving gNB. However, the effect of RLF parameters such as T310, N310 and N311 are not considered. 
Authors in \cite{ericsson_dynamic_bs} have emphasized the roles of link beam and access beams towards executing a handover in NR system. Therein, a reinforcement learning (RL) based approach to select the optimal neighboring gNB has been proposed accounting the RSRP measurements from access beams as state information. The goal of the RL based approach is to maximize the UE's throughput. In this work, the effect of RLFs and subsequent HFs has not been considered while designing the reward function for the RL model. Authors in \cite{ML-HF} propose a solution for handover management that optimizes handover parameters such as TTT, hysteresis (Hys) and A5 threshold to maximize edge user signal strength, load balancing and handover success rate simultaneously. \cite{9322339} proposes a smart Dual Connectivity triggering scheme for NR by which RLF caused by poor radio frequency conditions can be avoided. This scheme works by selecting the best B1 thresholds based on insights obtained from a Deep learning model to predict RLF. \cite{RLF_ML} uses an ML model that combines both Long Short Term Memory and Support Vector Machine to predict RLF. This ML model considers reference signal received quality, channel quality information and power head room as input. In \cite{7938389}, ML based approach has been used to group users into clusters based on their mobility patterns; and then adapt the TTT and Hys values. This work aims to optimize the data rate at the cell edge, as well as the rate of HFs. In \cite{HFPredict-2023}, an ML based method for HF prediction has been proposed based on some novel input features such as RSRP from serving/target cells along with interfering access networks. This ML model can predict HF with an accuracy of 93\%. In \cite{HPO}, a fuzzy logic based handover margin adaptation scheme has been proposed to optimize call dropping ratio (CDR) and number of handovers per successfully finished calls. \cite{Masri} introduces a method that leverages ML to learn local radio conditions and trigger handovers based on predicted radio environments. In \cite{Manalastas}, a data driven approach has been proposed to reduce inter-frequency handover failures by combining ML based transmit power tuning. {\bf It may be noted that these existing works \cite{refhandover,CHODeb, ericsson_dynamic_bs, ML-HF, 9322339, RLF_ML, 7938389, HFPredict-2023, HPO, Masri, Manalastas} to predict HF do not account for the effect of RLF adequately, even though it is one of the major reasons to cause HF \cite{lopez}.} 

Existing model based analyses of HF \cite{rlf_fail_prediction, Dynamic_UE_hyst_adjustment, HF_analysis_fading_hetnet} do not account for the aforementioned factors adequately. It may be noted that handover process is initiated if both the conditions \eqref{condition1} and \eqref{condition2} are TRUE for each and every sampling instances during $T_{prep}$ and $T_{exec}$. Now, assuming Rayleigh fading, the RSRP samples will follow exponential distribution. Moreover, these RSRP samples may be correlated as well. Similarly, RLF is also determined based on consecutive out-of-synch and in-synch indications which explicitly depends on the characteristics of the RSRP samples. Hence, the computation of HF probability based on the coincidence of RLF event during handover is subject to multivariate analysis of the underlying RSRP samples, which makes the model quite complex and intractable.  In such a prevailing situation, it is worthy to leverage Deep Q-Network (DQN) \cite{double_dqn} to model the RLF driven HF in terms of handover and RLF parameters. \emph{To the best of the authors' knowledge, this is the first power control algorithm towards minimizing RLF driven HF}. Our contributions are summarized as follows:

\begin{itemize}
\item We propose a DQN based power control algorithm, which takes RLF parameters (T310, N310), RSRP of serving and target gNB's and HF parameters ($T_{prep}$, $T_{exec}$, $O_{prep}$, $O_{exec}$) as input, and decides a possible increment of transmitting power of serving gNB to avoid RLF driven HF.  

\item The performance of the proposed DQN based power control algorithm has been investigated through extensive system level simulations. Results show that the traditional conditional handover mechanism, when equipped with the proposed DQN based power control, can significantly reduce RLF driven HF as compared to the conventional CHO \cite{CHODeb} as well as the RL based handover algorithm for 5G proposed in \cite{ericsson_dynamic_bs}.
\end{itemize}

The rest of the manuscript is organized as follows. In section \ref{DQN}, the proposed DQN based power control mechanism has been described. In section \ref{results}, the simulation results comparing between CHO, CHO + DRL and a RL based handover algorithm for 5G have been described. Finally, section \ref{con} concludes the work.  All the notations used in this study are summarized in  Table \ref{symbols}.

\begin{table}[htbp]
    \centering
    \caption{Symbol Table}
    \label{symbols}
    \begin{tabular}{|c|c|} \hline
        \textbf{Symbol} & \textbf{Meaning} \\ \hline
        RLF & Radio Link Failure \\ \hline
        HF & Handover Failure \\ \hline
        $T_{prep}$ & Time to preparation \\ \hline
        $O_{prep}$ & Preparation offset \\ \hline
        $T_{exec}$ & Time to execution \\ \hline
        $O_{exec}$ & Execution offset \\ \hline
        T310 & RLF timer \\ \hline
        N310 & \# out-of-synch indications to start T310 \\ \hline
        \end{tabular} \\ 
\end{table}

\section{Proposed DQN-based Power Control for Handover Failure Minimization} \label{DQN}
\subsection{DQN Framework for Dynamic Power Control}

In this work, we utilize \textbf{Deep Q-Network (DQN)} to minimize RLF driven HF. The considered RL setup involves an \textbf{agent} interacting with the \textbf{environment}, learning an optimal policy through trial and error. Definitions of states, actions, design of reward function and the DQN agent architecture is described as follows. 


\subsubsection{State Space Representation}
The efficacy of a DQN agent is critically dependent on its comprehensive perception of the environment. The state, $s_t$, at any given time step $t$, encapsulates key network conditions and User Equipment (UE) parameters essential for decision-making. The proposed DRL-based power control mechanism employs a 10-dimensional state vector, comprising:

\begin{itemize}
    \item \textbf{RSRP of serving gNB ($RSRP_{serv}$)}, i.e, the received signal strength from the currently serving gNB, measured in dBm, indicating current link quality;
    \item \textbf{RSRP of neighbouring gNB ($RSRP_{targ}$)}, i.e., the received signal strength from the strongest candidate gNB for handover, measured in dBm;
    \item \textbf{UE speed}, i.e., the UE's velocity in m/s, a critical factor influencing link stability and handover success;
    \item \textbf{Handover execution timer ($T_{exec}$)}, i.e., the configured duration (in ms) for the handover execution phase;
    \item \textbf{Handover preparation timer ($T_{prep}$)}, i.e., the configured duration (in ms) for the handover preparation phase.
    \item \textbf{Handover execution Offset ($O_{exec}$)}, i.e., the RSRP superiority margin (in dB) required for the target gNB to trigger the execution phase;
    \item \textbf{Handover Preparation Offset ($O_{prep}$)}, i.e., the RSRP superiority margin (dB) required for the target gNB to initiate the preparation phase;
    \item \textbf{RLF detection timer (T310)}, i.e., the configured duration (in ms) of the T310 timer, which, upon expiry after sustained out-of-sync conditions, leads to RLF declaration;
    \item \textbf{Out-of-sync counter threshold (N310)}, i.e., the number of consecutive out-of-sync indications required to initiate the T310 timer;
    \item \textbf{RLF threshold ($RSRP_{RLF}$)}, i.e., the signal strength threshold (in dBm) below which the UE is considered to be in out-of-synch conditions, potentially leading to an RLF event.
\end{itemize}
This rich feature set provides the agent with a detailed representation of the radio environment, active handover parameters, and RLF criteria.

\subsubsection{Action space definition}
The DQN network is configured with an output layer corresponding to two distinct actions, enabling the agent to adjust the serving gNB's transmission power. {\it The model is invoked whenever an out-of-sync indication is detected, signaling a potential risk of RLF.} At this point, the agent observes the current state $s_t$ and selects one of the following actions:

\begin{enumerate}
    \item \textbf{Action 0:} This action is selected when the agent predicts a higher likelihood of a RLF induced HF. To mitigate this, the agent requests an increase in the transmission power of the serving gNB by a predefined and discrete amount. Any requested power increase by the agent is subject to an absolute maximum gNB transmission power threshold namely $K$,  which the operational power level cannot surpass. The detailed operational characteristics and constraints of this power adjustment are elaborated in sub-section~\ref{subsec:simulation}.

    \item \textbf{Action 1:} This action is chosen when the agent estimates that the probability of an RLF induced HF is quiet low. As a result, the agent refrains from altering the serving gNB's transmission power, allowing the system to continue operating under existing configuration.
\end{enumerate}


\subsubsection{Reward function design}
\label{subsec:reward_fxn_refined}
The reward function is a critical component carefully designed to guide the DQN agent toward minimizing RLF-driven handover failures. Rewards and penalties are assigned based on specific events and their outcomes within the simulated environment and are evaluated every 20 ms. The rewards are structured to be relative, ensuring the agent learns to prioritize desirable behaviors over suboptimal ones.

Table~\ref{tab:reward_examples} lists seven prototypical scenarios, showing the agent’s decision, consequent   events and the resulting scalar rewards:

\begin{table*}[h!]
  \centering
  \caption{Illustrative Feedback and Rewards for Agent Decisions}
  \label{tab:reward_examples}
  \begin{tabular}{@{} l c c c c c c c r @{}}
    \toprule
    \textbf{Agent’s Decision} 
      & \textbf{Succ.\ HHO} 
      & \textbf{RLF} 
      & \textbf{Power Increase} 
      & \textbf{In‐synch} 
      & \textbf{Out‐synch} 
      & \textbf{Suppressed} 
      & \textbf{SINR Penalty} 
      & \multicolumn{1}{c}{\textbf{Reward}}
    \\ 
    \midrule
    Action\,0    & $\checkmark$ & $\times$     & $\times$         & $\times$       & $\times$        & $\times$        & $\times$           & $+\;15$   \\[6pt]
    Action\,0   & $\times$     & $\checkmark$ & $\checkmark$     & $\times$       & $\times$        & $\times$        & $\times$           & $-\;15$   \\[6pt]
    Action\,1    & $\times$     & $\checkmark$ & $\times$         & $\times$       & $\times$        & $\times$        & $\times$           & $-\;5$    \\[6pt]
    Action\,0    & $\times$     & $\times$     & $\checkmark$     & $\checkmark$   & $\times$        & $\times$        & $\times$           & $+\;5$    \\[6pt]
    Action\,1    & $\times$     & $\times$     & $\checkmark$     & $\times$       & $\checkmark$    & $\times$        & $\times$           & $-\;2$    \\[6pt]
    Action\,0    & $\times$     & $\times$     & $\times$ (suppressed) & $\times$   & $\times$        & $\checkmark$    & $\times$           & $-\;2$    \\[6pt]
    Action\,0    & $\times$     & $\times$     & $\checkmark$     & $\times$       & $\times$        & $\times$        & $\checkmark$       & $-\;\Delta\text{SINR}\times 300$ \\
    \bottomrule
  \end{tabular}
\end{table*}
\begin{itemize}
  \item  A reward of 15 units is given if handover succeeds due to Action 0 (Row 1).   
  \item A reward of –15 units is given if RLF occurs even though the agent took Action 0, thus discouraging such power increment (Row 2).  
  \item  A reward of -5 units is given if RLF occurs when no power increase was attempted, i.e., Action 1 is chosen by the agent. Such penalty discourages the agent to remain inactive when channel quality is poor (Row 3).
  \item  A reward of 5 units is given if an agent-initiated power increase (i.e., Action 0) recovers the link from out-of-sync phase, i.e., an in-sync followed by out-of-sync occurs (Row 4).
  \item  A reward of –2 units is given if Action 0 cannot restore the link from out-of-sync phase (Row 5). 
  \item A reward of –2 units is given if the agent Action 0 is suppressed, thus discouraging futile attempts during cooling window during which power adjustments are restricted (Row 6).  
  \item A reward of –\(\Delta\text{SINR}\times300\) is applied if the average SINR of neighboring UEs drops below a threshold within 40 ms of a power increase.   since, we evaluate every 20 ms, this penalty can be applied up to two times in 40 ms window (Row 7).  
\end{itemize}

\subsection{Deep Q-Network Agent Architecture}
Our DRL agent employs the Deep Q-Network (DQN) algorithm, a highly influential value-based reinforcement learning method. DQN learns an optimal action-value function, denoted as $Q^\pi(s,a)$, which estimates the expected cumulative discounted reward achievable by taking action $a$ in state $s$ and thereafter following a policy $\pi$. The optimal Q-function, $Q^*(s,a)$, is defined by the Bellman optimality equation  \cite{bellman1957dynamic}:

\begin{equation}
  Q^*(s,a) 
  \;=\; 
  \mathbb{E}_{\,s' \sim P(\,\cdot\,\mid s,a)}\Bigl[\,
    r(s,a,s') 
    \;+\;\gamma \max_{a'} Q^*(s',a')\Bigr]
\end{equation}

where $P(s' \mid s,a)$ is the probability of transitioning to state $s'$ from $(s,a)$, 
$r(s,a,s')$ is the immediate reward received upon that transition, 
 $\gamma \in [0,1]$ is the discount factor that balances immediate versus future rewards, and in our implementation, is set to $0.95$.

\subsubsection{Neural Network Model}
We approximate the action-value function $Q(s,a)$ using a deep neural network, which serves as a core component of our DQN agent. This network takes the current state vector $s_t$ as input and outputs the estimated Q-values for each of the defined discrete actions. Our implemented architecture features:
\begin{itemize}
    \item Input layer: An input layer compatible with the 10-dimensional normalized state vector.
    \item Hidden layers: Three fully connected (dense) hidden layers. Each layer utilizes the Rectified Linear Unit (ReLU) activation function\cite{relu}, which introduces non-linearity crucial for approximating complex value functions. These hidden layers are configured with 64 neurons each.
    \item Output layer: A fully connected linear output layer with two neurons, corresponding to the two actions in the agent's action space. The linear activation allows the Q-values to take on any real value.
\end{itemize}
This neural network structure enables the agent to learn intricate mappings from states to action-values.


\subsubsection{Training enhancements}
To ensure robust and efficient convergence during the learning process, the DQN agent incorporates well known techniques such as experience reply, employing Double DQN and Huber loss function, and $\epsilon$‐greedy exploration \cite{double_dqn, huber-loss, greedy-exploration, exploration}.  

\begin{itemize}
    \item Experience replay: Past experiences are stored as transitions 
  \[
    (s_t,\,a_t,\,r_t,\,s_{t+1},\,done_t),
  \]
      where 
      \begin{itemize}
        \item $s_t$ is the state at time $t$, 
        \item $a_t$ is the action taken at time $t$, 
        \item $r_t = r(s_t,a_t,s_{t+1})$ is the immediate reward, 
        \item $s_{t+1}$ is the successor state, and 
        \item $done_t \in \{0,1\}$ is a Boolean flag indicating whether $s_{t+1}$ is terminal ($done_t=1$ means episode ends at $t+1$, else $done_t=0$).
      \end{itemize}
    During training, mini-batches of these transitions are randomly sampled from the buffer. This practice de-correlates the data used for updates, breaking temporal dependencies and smoothing the learning process by averaging over a diverse set of past experiences.  
    
    \item Target network and Double DQN: To stabilize learning, we use Double DQN, first introduced in \cite{double_dqn}, where two separate neural networks are employed: a \textit{policy network} ($Q_{\theta}$), which is actively updated and used for action selection, and a \textit{target network} ($Q_{\theta'}$), which provides the target Q-values for the Bellman updates. The target network's weights ($\theta'$) are periodically synchronized with the policy network's weights ($\theta' \leftarrow \theta$), creating a more stable learning target. Furthermore, the \textbf{Double DQN} refinement is utilized. This technique mitigates the overestimation bias common in standard Q-learning by decoupling the selection of the best next action from its value estimation. The policy network determines the optimal next action ($a^*_{t+1} = \arg\max_{a'} Q_{\theta}(s_{t+1}, a')$), but the target network evaluates its Q-value ($Q_{\theta'}(s_{t+1}, a^*_{t+1})$). The  Double DQN target at time t is thus:
\begin{align}
    y_t^{\text{DDQN}} = r_t + \gamma (1 - done_t) \, Q_{\theta'}\big(&s_{t+1}, \nonumber \\
    &\arg\max_{a'} Q_{\theta}(s_{t+1}, a') \big)
\end{align}

\item $\epsilon$‐greedy exploration: 
To balance between exploration of new actions and exploitation of known optimal actions, we use $\epsilon$‐greedy strategy \cite{greedy-exploration}. At each decision step, with probability $\epsilon$ the agent selects a random action; otherwise (with probability $1 - \epsilon$) it chooses the action that maximizes the current estimated Q‐value. The exploration rate $\epsilon$ is annealed exponentially from an initial value of $\epsilon_{0} = 1.0$ down to a minimum of $\epsilon_{\mathrm{min}} = 0.01$ over $\epsilon_{\text{decay}} = 5000$ steps:
\[
  \epsilon_{t} \;=\; 
    \epsilon_{\mathrm{min}} \;+\; 
    \bigl(\epsilon_{0} - \epsilon_{\mathrm{min}}\bigr)\,e^{-\,t / \epsilon_{\text{decay}}}\,,
\]
where $t$ is the global training step (i.e.\ the total number of action‐selection steps completed so far).        

\item \textbf{Loss function optimization:} 
The policy network is trained by minimizing the discrepancy between its predicted Q‐values and the target Q‐values defined by the Double DQN update. To this end, we employ the Huber loss (also known as Smooth L1 loss) \cite{huber-loss}:
\[
  L_{\delta}(\mathit{err}) =
    \begin{cases}
      \displaystyle \tfrac{1}{2}\,\mathit{err}^2, 
        & \text{if } |\mathit{err}| \le \delta, \\[0.8em]
      \displaystyle \delta \bigl(|\mathit{err}| - \tfrac{1}{2}\,\delta\bigr), 
        & \text{if } |\mathit{err}| > \delta,
    \end{cases}
\]
where 
\[
  \mathit{err} \;=\; y_{t}^{\mathrm{DDQN}} - Q_{\theta}(s_{t}, a_{t}),
  \quad
  \delta = 1.
\]
Here, $y_{t}^{\mathrm{DDQN}}$ is the Double DQN target as defined previously. 
Huber loss is selected for its robustness to outliers compared to mean squared error, 
while still providing smooth gradients near the optimum. 
We optimize using the Adam algorithm (an adaptive‐learning‐rate method) 
and apply gradient clipping with a global‐norm threshold of 10 to prevent excessively large gradients. 

\end{itemize}

The training involves the DRL agent engaging in numerous episodes of interaction with the simulated 5G environment. Within each episode, the agent sequentially observes states, selects actions based on its current policy, receives corresponding rewards, and stores these experiences. The policy network is updated using mini-batches of experiences sampled from the replay buffer. The ultimate aim is to converge to an optimal policy $\pi^*$ that maximizes the expected cumulative discounted reward, thereby realizing an intelligent and adaptive power control strategy for the effective mitigation of RLFs and subsequent handover failures.
 
The overall DRL training loop, integrating these components, is formally presented in Algorithm~\ref{alg:dql_training}. The variables used in the algorithm are mentioned in ~\ref{tab:algorithm_variables} for clarity:

\begin{table}[htbp]
\centering
\caption{Definition of Variables Used in Algorithm~\ref{alg:dql_training}}
\label{tab:algorithm_variables}
\begin{tabular}{ll}
\toprule
\textbf{Variable} & \textbf{Definition} \\
\midrule
$n_{\text{state}}$ & Dimension of the state space \\
$n_{\text{action}}$ & Number of discrete actions available \\
$N_{\text{buffer}}$ & Capacity of the experience replay buffer \\
$B$ & Mini‐batch size sampled from the buffer \\
$\gamma$ & Discount factor for future rewards \\
$N_{\text{start}}$ & Minimum steps before training begins \\
$C_{\text{target}}$ & Frequency (in steps) of target network updates \\
$\epsilon_{\text{start}}, \epsilon_{\text{end}}$ & Initial and final exploration probabilities \\
$\tau_{\epsilon}$ & Decay constant for $\epsilon$‐greedy annealing \\
$G_{\max}$ & Maximum gradient norm for clipping \\
$M_{\text{episodes}}$ & Total number of training episodes \\
$T_{\text{steps}}$ & Maximum time steps per episode \\
\bottomrule
\end{tabular}
\end{table}


\begin{algorithm}
\caption{Deep Q‐Network based Power Control }
\label{alg:dql_training}
\begin{algorithmic}[1]
    \REQUIRE 
    $n_{\text{state}},\,n_{\text{action}},\,N_{\text{buffer}},\,B,\,\gamma,\,N_{\text{start}},\,C_{\text{target}},$ \\
    \hspace{1.5em}$\epsilon_{\text{start}},\,\epsilon_{\text{end}},\,\tau_{\epsilon},\,G_{\max},\,M_{\text{episodes}},\,T_{\text{steps}}$
    \STATE Initialize replay buffer $D$ (capacity $N_{\text{buffer}}$)
    \STATE Initialize policy network $Q_{\theta}$ and set target network $Q_{\theta'} \gets Q_{\theta}$
    \STATE $\text{steps\_done} \gets 0,\;\epsilon \gets \epsilon_{\text{start}}$
    \FOR{episode $=1$ to $M_{\text{episodes}}$}
        \STATE $s \gets$ initial state (e.g., emulator reset)
        \FOR{$t = 1$ to $T_{\text{steps}}$}
            \STATE $\text{steps\_done} \gets \text{steps\_done} + 1$
            \STATE $\epsilon \gets \epsilon_{\text{end}} \;+\; (\epsilon_{\text{start}} - \epsilon_{\text{end}})\,\exp\bigl(-\,\text{steps\_done}/\tau_{\epsilon}\bigr)$
            \IF{rand() $< \epsilon$}
                \STATE Select random action $a$
            \ELSE
                \STATE $a \gets \arg\max_{a} Q_{\theta}(s,a)$
            \ENDIF
            \STATE Execute $a$, observe $(r,\,s',\,\text{done})$, store $(s,a,r,s',\text{done})$ in $D$
            \IF{$\text{steps\_done} > N_{\text{start}}$ \AND $|D|\ge B$}
                \STATE Sample minibatch $\{(s_j,\,a_j,\,r_j,\,s'_j,\,d_j)\}_{j=1}^B$ from $D$
                \STATE $Q(s_j, a_j) \gets Q_{\theta}(s_j)[\,a_j\,]$ for each $j$
                \STATE $Q_{\text{target}} \gets Q_{\theta'}\big(s'_j,\,\arg\max_{a'} Q_{\theta}(s'_j, a')\big)$ 
                \STATE $\hat{Q}(s_j, a_j) \gets r_j + (1 - d_j)\,\gamma\,Q_{\text{target}}$  for each j
                \STATE $\mathcal{L} \gets \mathrm{smooth\_}L_{1}\big( Q(s_j, a_j),\,\hat{Q}(s_j, a_j)\big)$
                \STATE Backpropagate $\mathcal{L}$, clip gradients to norm $G_{\max}$, update $\theta$
            \ENDIF
            \IF{$\text{steps\_done} \bmod C_{\text{target}} = 0$}
                \STATE $\theta' \gets \theta$
            \ENDIF
            \STATE $s \gets s'$
            \IF{$\text{done}$}
                \STATE \textbf{break}
            \ENDIF
        \ENDFOR
    \ENDFOR
\end{algorithmic}
\end{algorithm}

\subsection{System aspects}
The proposed agent for each UE is implemented in Radio resource control (RRC) layer at the gNB. The RRC layer also controls the handover \cite{rrc_3gpp}. All RLF parameter values (N310, N311, T310 and $RSRP_{RLF}$) are configured in RRC layer of the gNB and broadcasted to the UE via the dedicated RRC reconfiguration messages. The UE detects RLF and sends it via the RRC \emph{UEInformationResponse}, or during the reestablishment of communication. The gNB also stores the CHO parameters and sends the values to the UE via an RRC configuration message. Furthermore, the RRC layer is responsible for configuring the {\it base power} for \texttt{broadcast PDSCH power offsets}, and defines the limits and constraints to compute the transmit power per UE. The trained model can create instances for each UE going through the handover process.

\section{Results and Discussions} \label{results}
To evaluate the effectiveness of our proposed DQN-based  power control mechanism, we developed a simulation framework aligned with 5G NR standards. The framework simulates UE mobility, handovers, path loss, fading and obstacle‐induced link degradation. We compare the performance of our proposed algorithm with the RL based handover mechanism in \cite{ericsson_dynamic_bs}, in terms of RLF and RLF induced HF. In \cite{ericsson_dynamic_bs}, an RL agent is trained to choose the optimal neighboring gNB during a handover procedure. The RL agent considers the RSRP measurements from the access beams as state information. Therein, the reward for each action is the difference between the received power through the link beams of the previously serving gNB and the newly chosen gNB. In the next subsection, we describe the simulation set-up.

\subsection{Simulation setup}
\label{subsec:simulation}
Python is chosen for simulation due to its simplicity, robustness and extensive libraries supporting numerical computation, RL and wireless network modeling. Libraries such as \textbf{NumPy} and \textbf{Matplotlib} were used for signal processing and visualization, while \textbf{PyTorch} enabled the DQN-based learning module.
The full simulation code and scripts used to generate the results are publicly available on GitHub \cite{kartheek2025chodqn}.

Signal attenuation (in dB) over distance is captured via the log‑distance path loss model: 
\begin{equation}
  PL(d) = 10\,\alpha\,\log_{10}\left(\frac{d}{d_0}\right);
  \quad
  \alpha = 2.8,
  \quad
  d_0 = 1\text{\,m}.
\end{equation}
Here, $d$ is the distance (in meters) between UE and gNB. The exponent $\alpha=2.8$ reflects measured urban/suburban propagation \cite{paper_ple}, and $d_0$ ($=1$m) normalizes the loss at close range. As $d$ increases, $PL(d)$ grows logarithmically, ensuring that distant cells exert progressively less influence on the received signal.

Small‑scale multipath ($F$) is incorporated by sampling a complex Gaussian coefficient \(h=x+jy\), \(x,y\sim\mathcal{N}(0,1)\) and applying it to the path‐loss gain $\left(d_0/d \right)$ as follows:
\[
  F = 10\log_{10}\Bigl\lvert \frac{d_0}{d}\,h\Bigr\rvert^2.
\]
This term models the rapid fluctuations in instantaneous power that the DRL agent must learn to counteract.

Static and dynamic obstructions (buildings, foliage, vehicles) further degrade links by blocking or scattering energy. We employ the following empirical LoS probability model\cite{Lod-prob}:
\begin{equation}
  p_{\text{LOS}}(d) = \min\!\Bigl(\tfrac{20}{d},1\Bigr)\,(1 - \epsilon^{d/39}) + \epsilon^{d/39},
  \quad
  \epsilon = 0.8.
\end{equation}
This expression blends a distance‑based cutoff ($20/d$) with an exponential decay ($\epsilon^{d/39}$) that captures urban density. When LoS exists, signal attenuation is small; otherwise the link suffers an additional loss. Accordingly, the received power (in dB) at the UE has been computed as:
\begin{equation}
  P_{\text{recv}} = P_t - PL(d) + F + 10\log_{10}\bigl(p_{\text{LOS}}(d)\bigr).
\end{equation}
Thus, $p_{\text{LOS}}(d)$ acts as a continuous attenuator, smoothly transitioning between clear and obstructed conditions.

\begin{figure}[ht]
  \centering
  \includegraphics[width=0.9\linewidth]{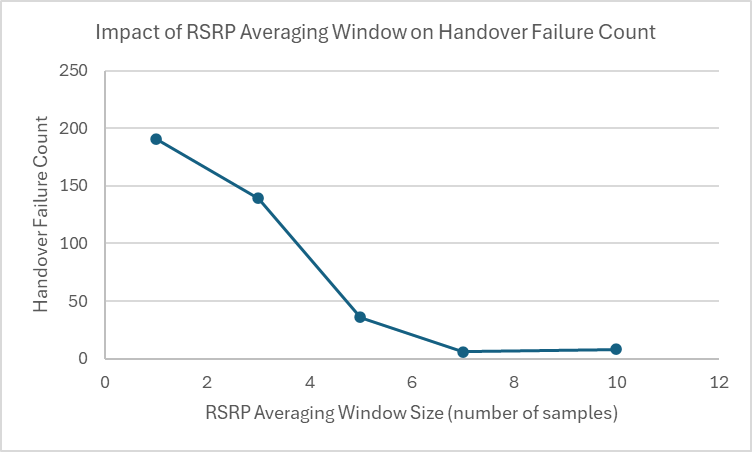}
  \caption{Effect of RSRP averaging window size \(N\) on HF Count.}
  \label{fig:hof_vs_avg}
\end{figure}

To reduce short‐term fluctuations in the measured RSRP and improve the stability of the handover decision process, we compute a moving average over the last \(N\) samples:
\begin{equation}
  \overline{RSRP}_k = \frac{1}{N} \sum_{i=k-N+1}^{k} RSRP_i.
\end{equation}
In our experiments, we varied \(N\) (the averaging window size) and recorded the resulting HF count (depicted in Fig. \ref{fig:hof_vs_avg}). We evaluated HF for several averaging window sizes \(N\in\{1,3,5,7,10\}\).  The result shows that although larger \(N\) yields a smoother RSRP trace (and hence fewer spurious handovers), excessively large windows introduce delay in reacting to genuine signal drops. Based on this trade‐off, we selected \(N=5\) as the final averaging window, which provides a stable yet responsive RSRP estimate.

\begin{table}[ht]
  \caption{Default simulation parameters}
  \label{tab:default_params}
  \centering
  \begin{tabular}{l l}
    \hline
    \textbf{Parameter}                        & \textbf{Value} \\
    \hline
    \multicolumn{2}{l}{\textit{Simulation Parameters}} \\
    \hline
    Number of eNB                             & 15 \\
    Service area                              & 3000 x 500\,${m^2}$ \\
    Frequency band                            & 25\,GHz \\
    Max bandwidth                             & 100\,MHz \\
    gNB transmit power (initial / max)        & 33\,dBm / 38.5\,dBm \\
    UE speed                                  & 40\,km/h \\
    Path loss exponent ($\alpha$)             & 2.8 \\
    Reference distance ($d_0$)                & 1\,m \\
    Fading/shadowing                        & Enabled \\
    Obstacle-induced LOS model                & Enabled ($\epsilon=0.8$) \\
    RLF threshold ($S_{\text{RLF}}$)          & -67.5\,dBm \\
    N310 threshold                            & 6 \\
    T310 timer                                & 1000\,ms \\
    O$_{prep}$                                & 1\,dB \\
    O$_{exec}$                                & 6\,dB \\
    T$_{prep}$                                & 100\,ms \\
    T$_{exec}$                                & 80\,ms \\
    \hline
    \multicolumn{2}{l}{\textit{Training Parameters}} \\
    \hline
    Discount factor ($\gamma$)                & 0.95 \\
    Epsilon (start/end)                     & 1.0 / 0.01 \\
    Epsilon decay constant ($\tau_{\epsilon}$) & 5000 \\
    Replay buffer size ($N_{\text{buffer}}$)  & 10000 \\
    Minibatch size ($B$)                      & 64 \\
    Training start threshold ($N_{\text{start}}$) & 50 steps \\
    Target network update frequency ($C_{\text{target}}$) & 100 steps \\
    Max gradient norm ($G_{\max}$)            & 10 \\
    Episodes ($M_{\text{episodes}}$)          & 2000 \\
    \hline
  \end{tabular}
\end{table}

The simulation environment incorporates a detailed model for dynamic adjustments to the serving gNB's transmission power when triggered by the DRL agent (Action 0). This model includes the following key characteristics:
\begin{itemize}
    \item Power increment: The transmission power is increased by a predefined, discrete value (2000 mW) when Action 0 is chosen by the agent. This increment is a configurable simulation parameter. Here $K$ has been set to  38.5 dBm.
    
    \item Temporary boost and automatic reversion: An initiated power increase due to Action 0 of the agent is not permanent. It acts as a temporary boost for a specific, configurable duration. After this period, the gNB's transmission power automatically reverts to its operational level prior to the boost.
    \item Cooldown protocol: To ensure network stability and prevent overly rapid or oscillating power adjustments, a \emph{cooldown protocol} is implemented. Following a power increase and its subsequent reversion, or if an increase attempt is made while a boost is still active, further power increase commands may be temporarily disallowed or penalized. This protocol manages the frequency of power boosts.
\end{itemize}

In our simulation, the out-of-sync ($S_{RLF}$) and in-sync ($Q_{in}$) threshold for RLF detection has been considered to be equal. Default simulation parameters are depicted in Table \ref{tab:default_params}.

\subsection{Training the DQN Algorithm}
The DQN agent is trained in a simplified but representative 2-gNB setup across \textbf{2000 episodes}, using the reward structure defined in Section~\ref{subsec:reward_fxn_refined}. Unlike a fixed-parameter training regime, we intentionally vary several critical handover related parameters during training to promote generalization. These include offsets ($O_{prep}$, $O_{exec}$), timers ($T_{prep}$, $T_{exec}$, T310), N310 thresholds and RLF thresholds, as detailed in the parameter options configuration. This exposes the agent to a broad range of handover conditions and failure scenarios.

To simulate diverse mobility patterns, the UE speed is varied per episode using a random scaling factor (e.g., between $0.8$ and $1.2$) applied to a base range of $35$–$45$ km/h. The UE follows a straight-line trajectory from gNB 1 to gNB 2 over $10,000$ ms, allowing multiple handover opportunities in each episode. Further, gNB 1 is randomly placed at $\bigl(x_1,y_1\bigr)$ with $x_1\sim\mathcal{U}[2,100]$ m and $y_1\sim\mathcal{U}[230,240]$ m, and its transmit power varies between $33$–$40$ dBm. gNB 2 is randomly placed at $\bigl(x_2,y_2\bigr)$ with $x_2\sim\mathcal{U}[200,350]$ m and $y_2\sim\mathcal{U}[260,270]$ m, with fixed transmit power of $33$ dBm.

By systematically varying handover parameters while keeping other simulation aspects fixed (as listed in Table~\ref{tab:default_params}), the agent learns to generalize across different CHO settings. Over 2000 episodes, we track cumulative reward, loss convergence, and the distribution of handover and RLF events to ensure convergence toward a robust and adaptable policy.

\subsection{Testing the DQN algorithm}
\begin{figure}[ht]
  \centering
  \includegraphics[width=\linewidth]{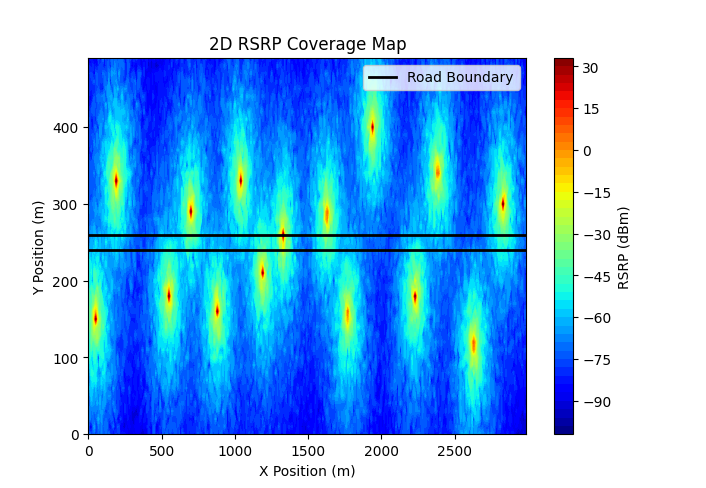}
  \caption{Heatmap of RSRP across the 2D gNB deployment. The UE trajectory (from \(x=0\) to \(3000\) m at \(y\approx500\) m) is overlaid.}
  \label{fig:power_heatmap}
\end{figure}
After convergence, the trained model is evaluated in a more complex scenario with 15 gNBs placed across a 2D plane. Table \ref{tab:enb_positions} lists the (x,y) coordinates and intended purpose of each gNB, while Table~\ref{tab:pairwise_distances} reports the pairwise distances between consecutive stations. These irregular spacings (from 148.7 m up to 390.0 m) ensure that handovers occur under diverse signal‑overlap and gap scenarios. The co-ordinates in Table \ref{tab:enb_positions} have been chosen to create a variety of coverage conditions ranging from well‑served zones to engineered \emph{black spots} and weak‑transition areas to rigorously evaluate the generalization capability of the proposed DQN agent. To emulate realistic variations, the UE's nominal speed of 40\,km/h is also scaled by a random factor between 0.8 and 1.2 at every step during testing. During each test run, a UE traverses in a straight line from \(x=0\)m to \(x=3000\)m at a constant step \(y\approx250\) m, stimulating handover events across the entire topology (depicted in Figure \ref{fig:power_heatmap}). This phase assesses the policy's ability to generalize under unseen configurations with varied offset thresholds and timer values. In both training and testing phases, the DQN agent adapts transmit power to minimize RLFs while preserving efficient handovers.

\begin{figure*}[!htb]
  \centering
  \subfloat[RLF vs. N310]{\includegraphics[width=5cm, height=5cm]{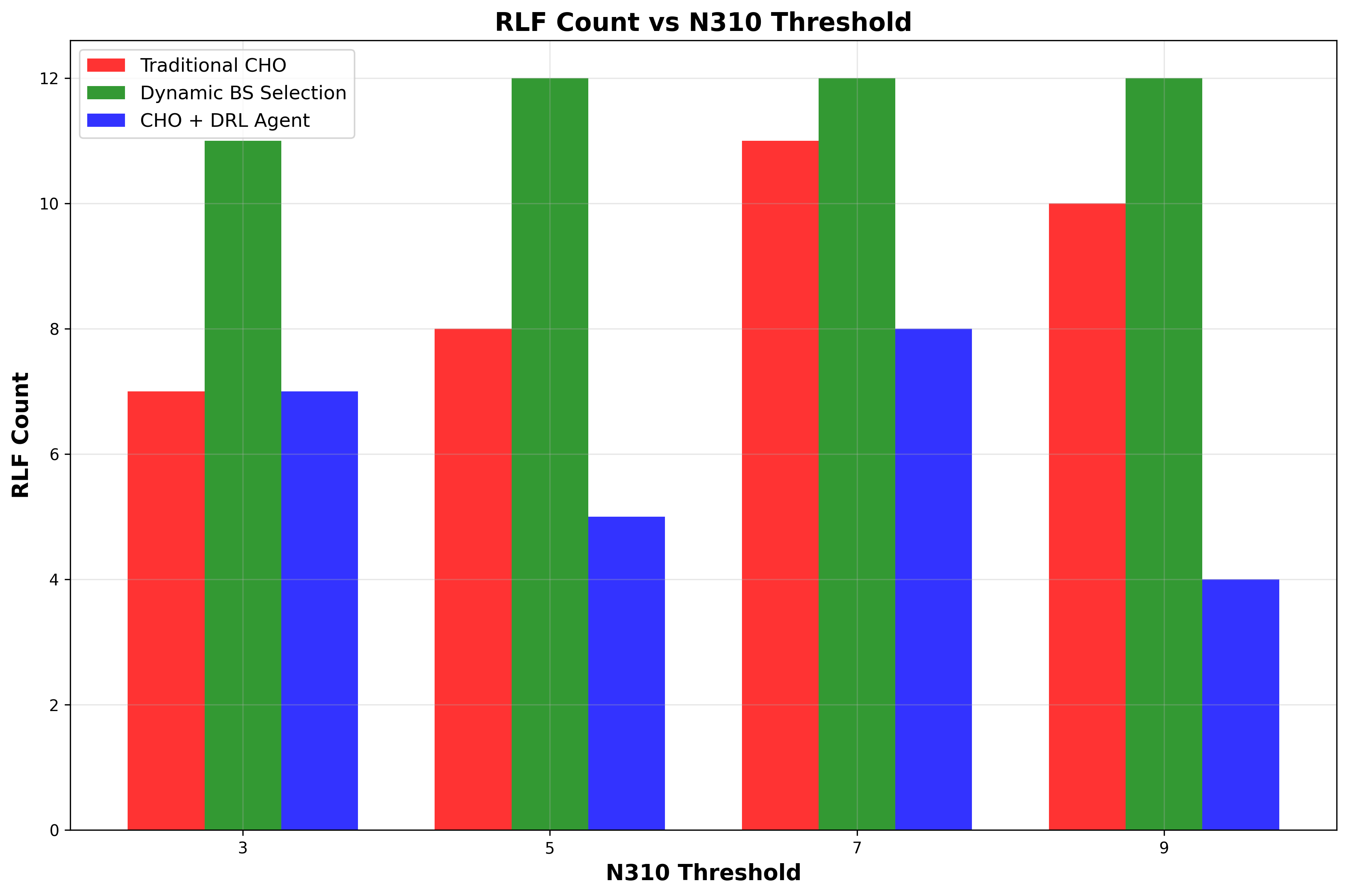}}
  \subfloat[RLF vs. $O_{exec}$]{\includegraphics[width=5.5cm, height=5cm]{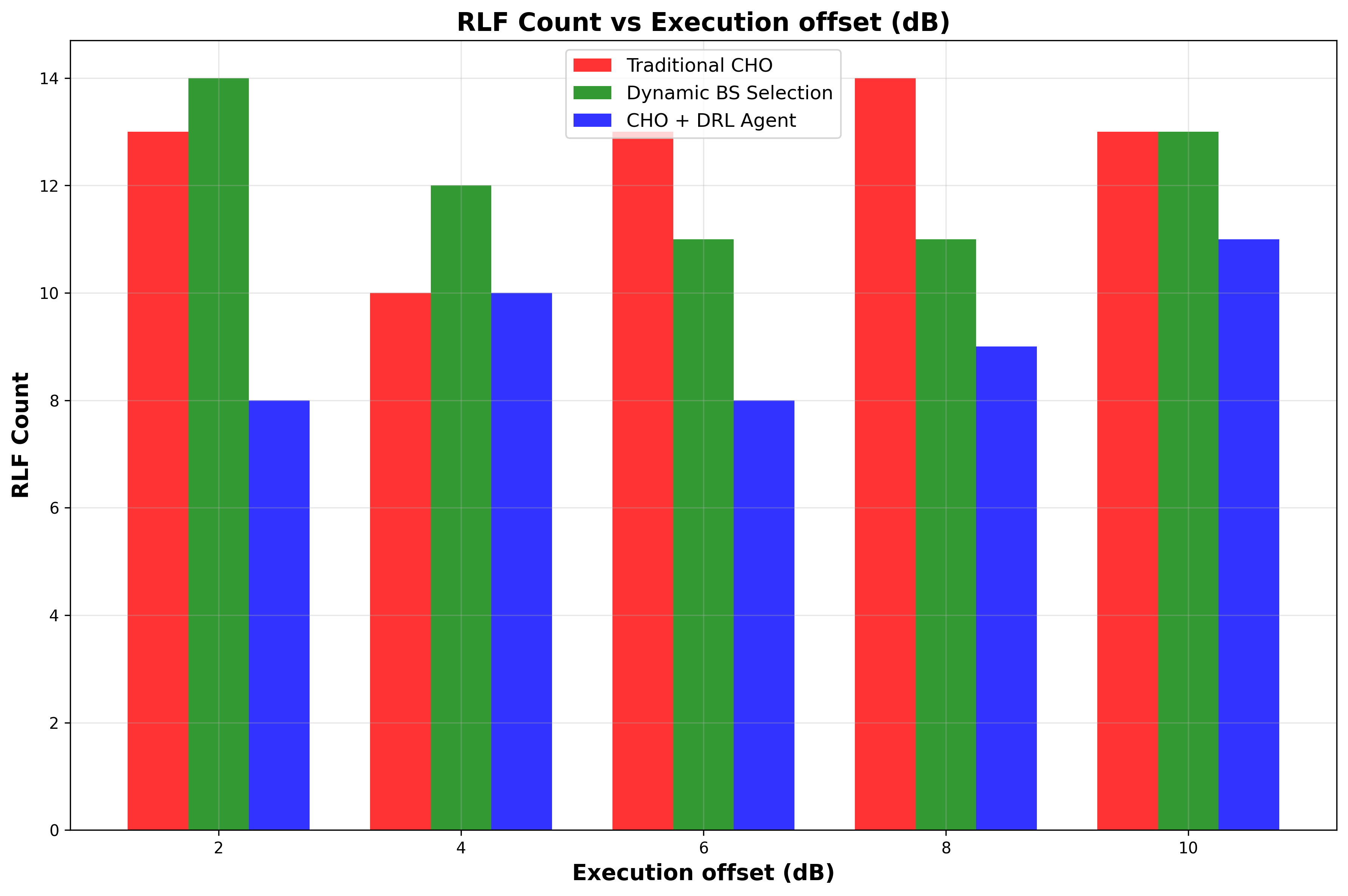}}
  \subfloat[RLF vs. $O_{prep}$]{\includegraphics[width=5.5cm, height=5cm]{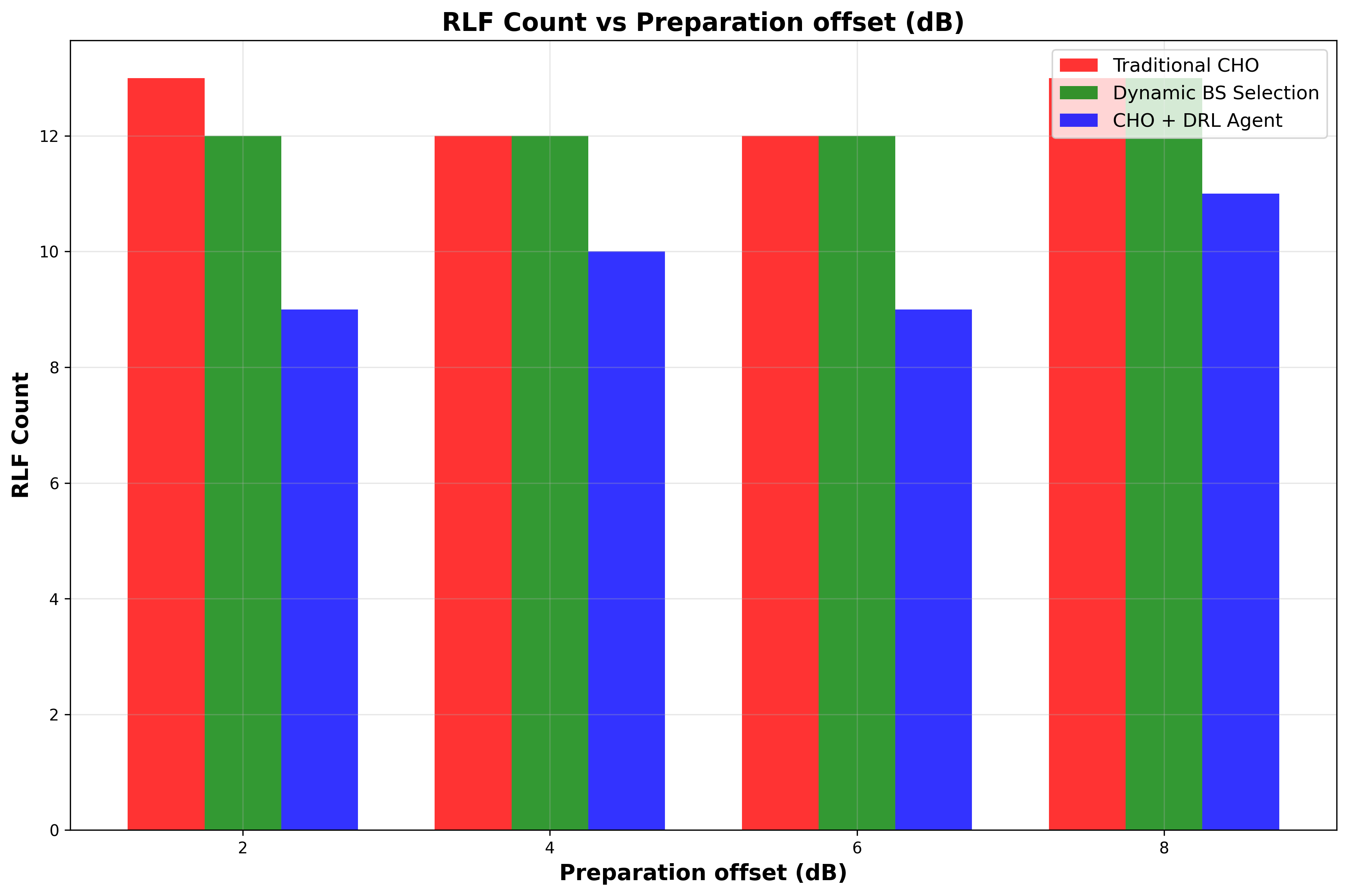}} \\
  \subfloat[RLF vs. T310]{\includegraphics[width=5cm, height=5cm]{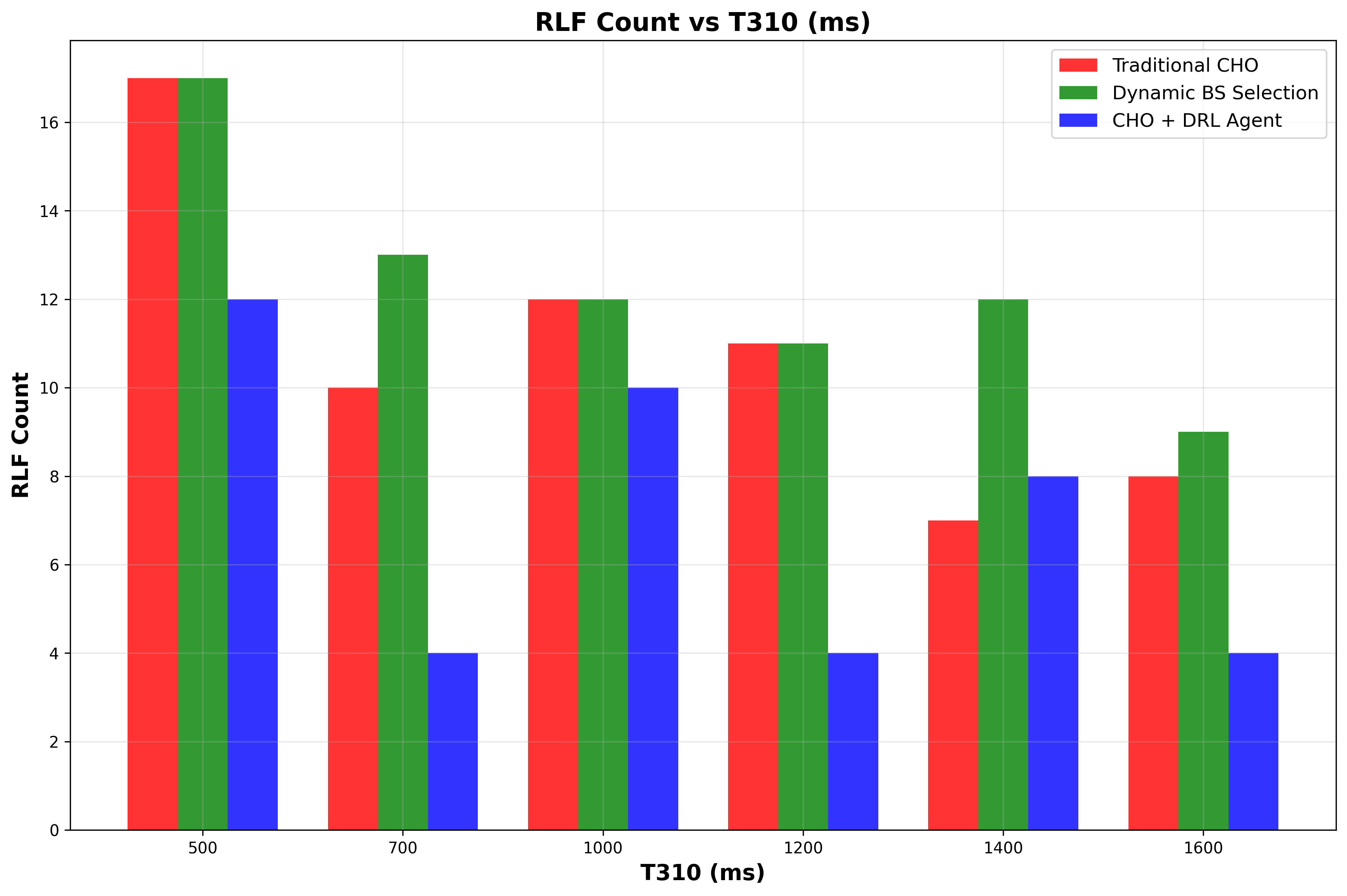}}
  \subfloat[RLF vs. $T_{exec}$]{\includegraphics[width=5.5cm, height=5cm]{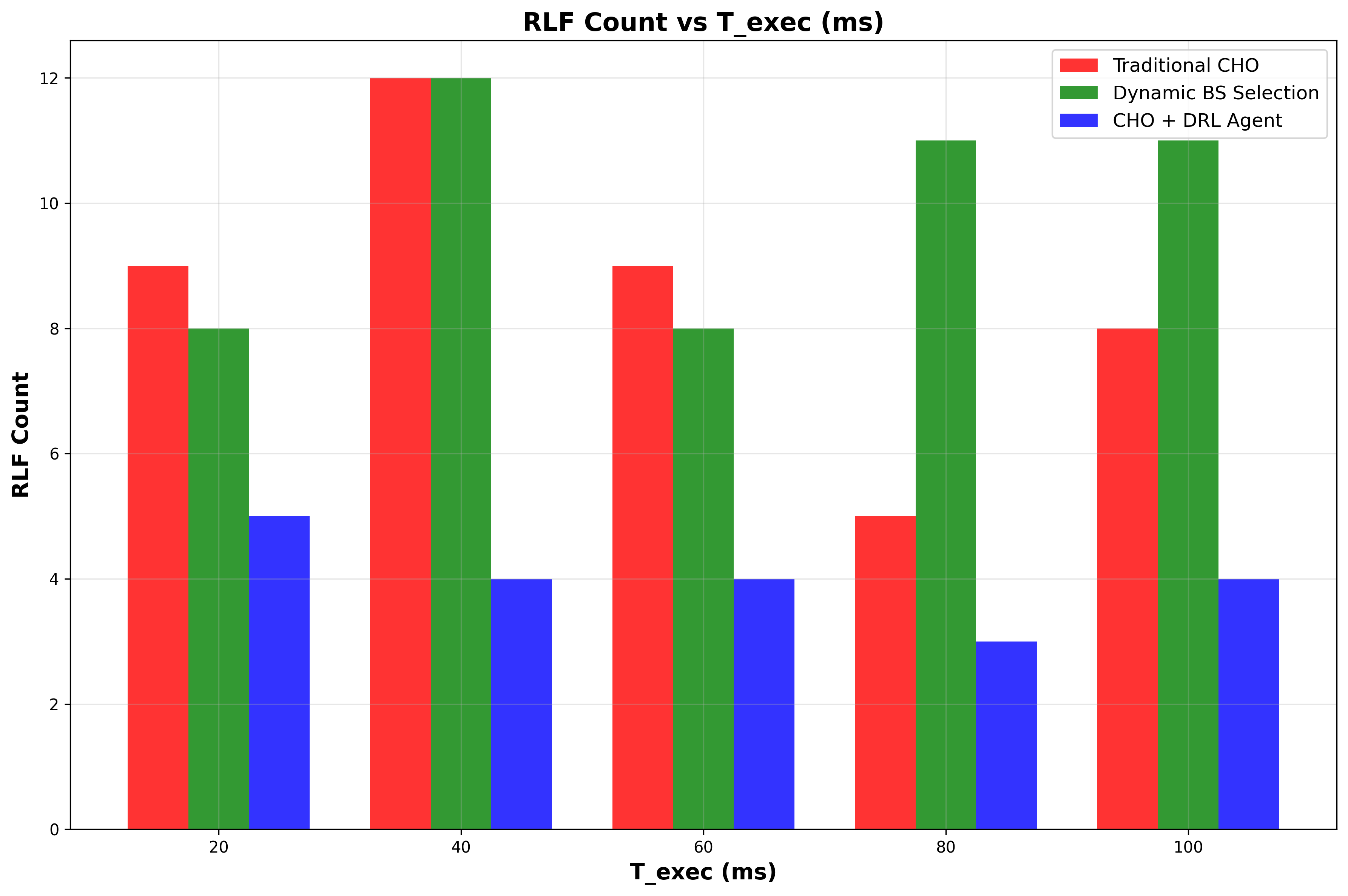}}
  \subfloat[RLF vs. $T_{prep}$]{\includegraphics[width=5.5cm, height=5cm]{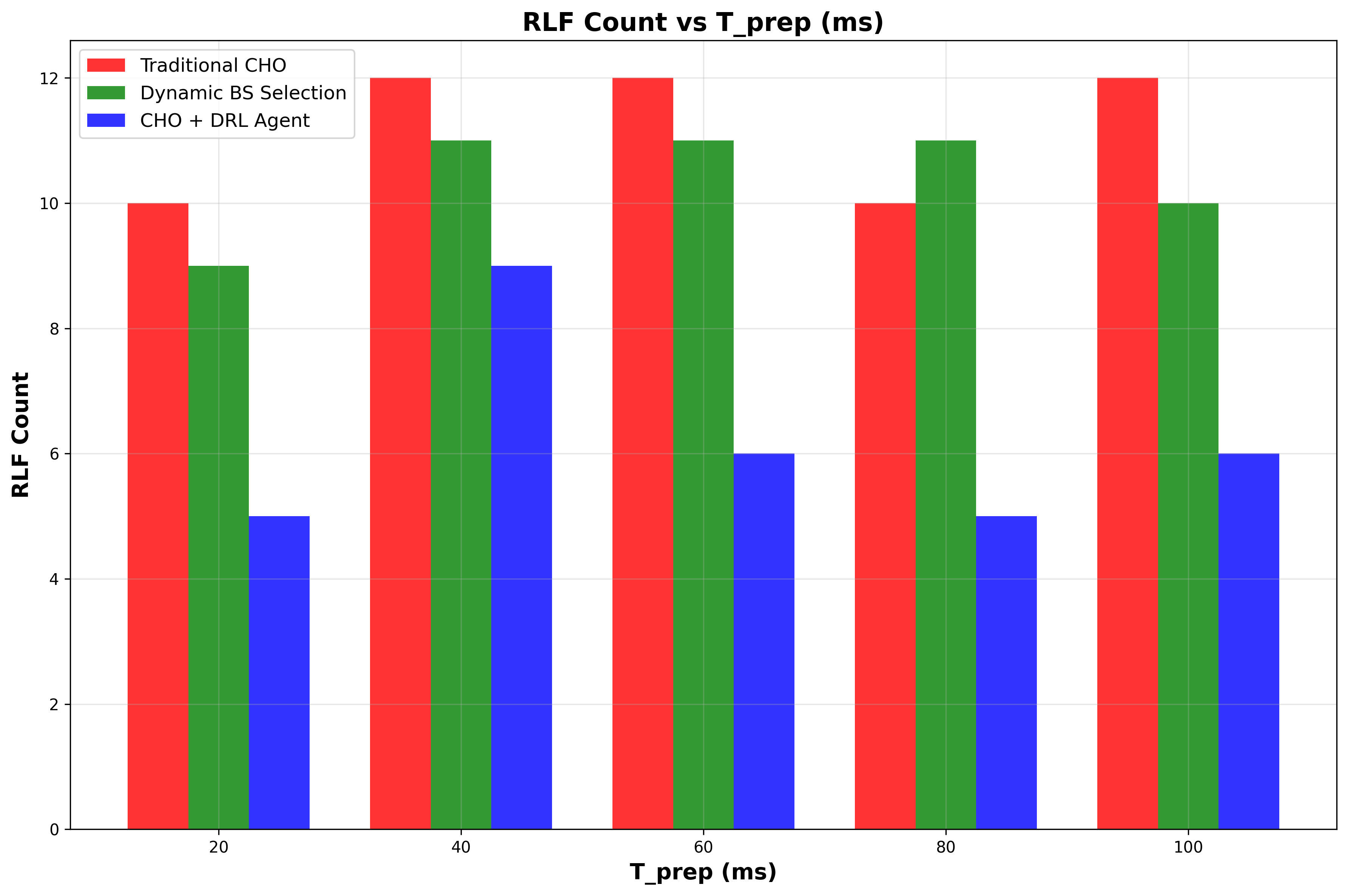}}
  \caption{RLF counts for CHO and CHO+DRL across different parameter sweeps}
  \label{fig:rlf_comparison}
\end{figure*}

\begin{figure*}[htb]
  \centering
  \subfloat[HF Count vs. N310]{\includegraphics[width=5.5cm, height=5cm]{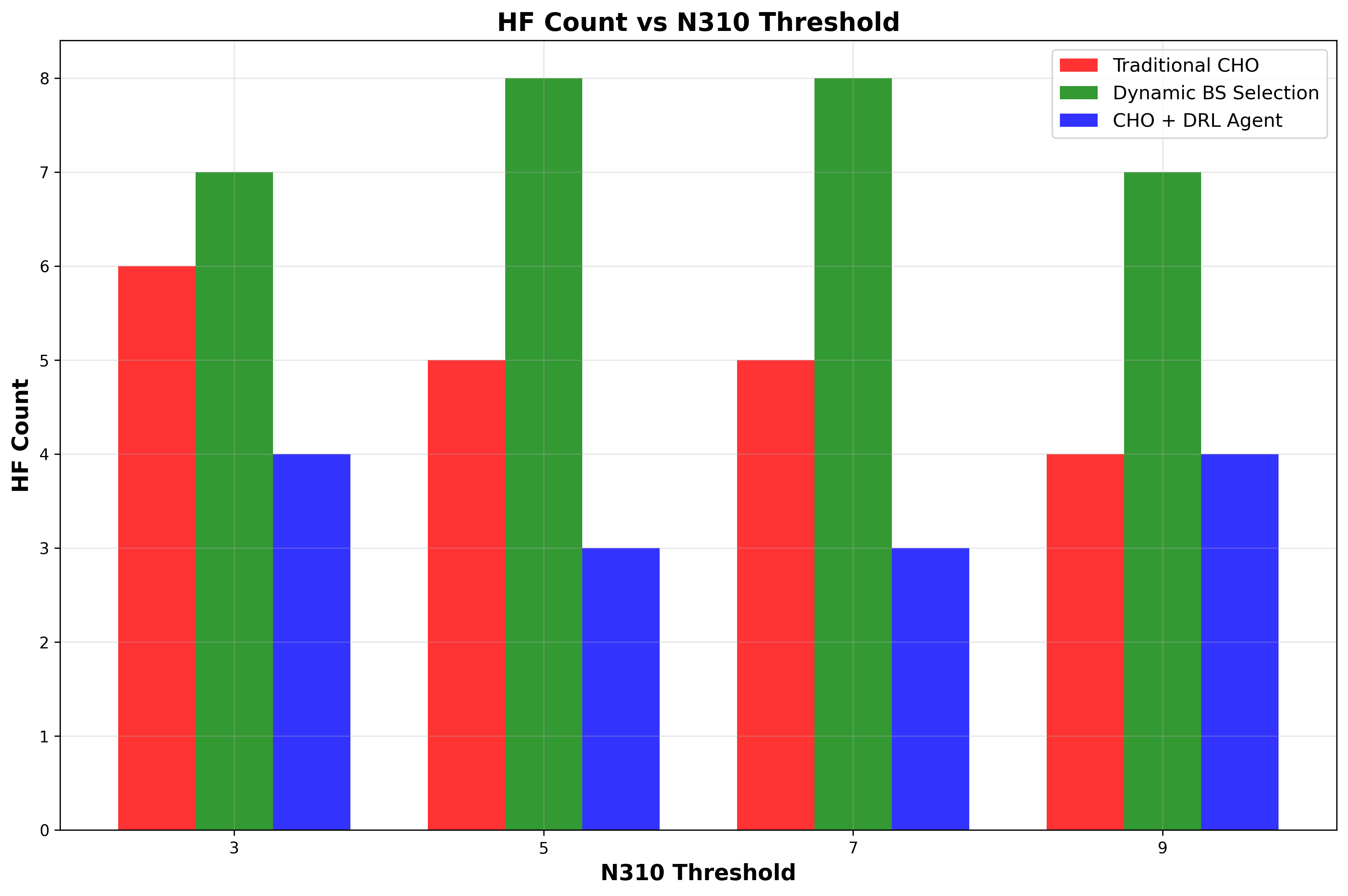}}
  \subfloat[HF Count vs. $O_{exec}$]{\includegraphics[width=5.5cm, height=5cm]{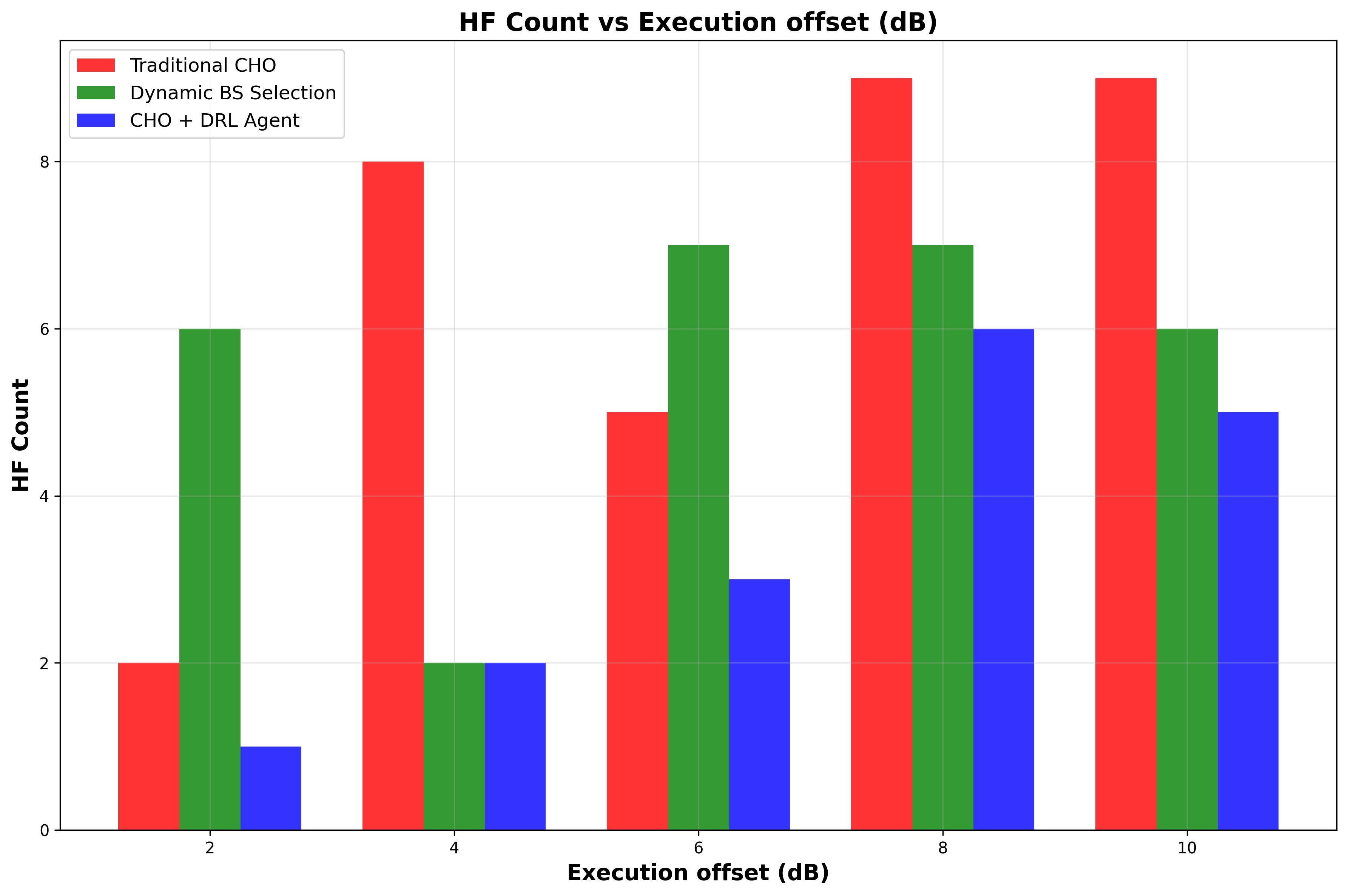}}
  \subfloat[HF Count vs. $O_{prep}$]{\includegraphics[width=5.5cm, height=5cm]{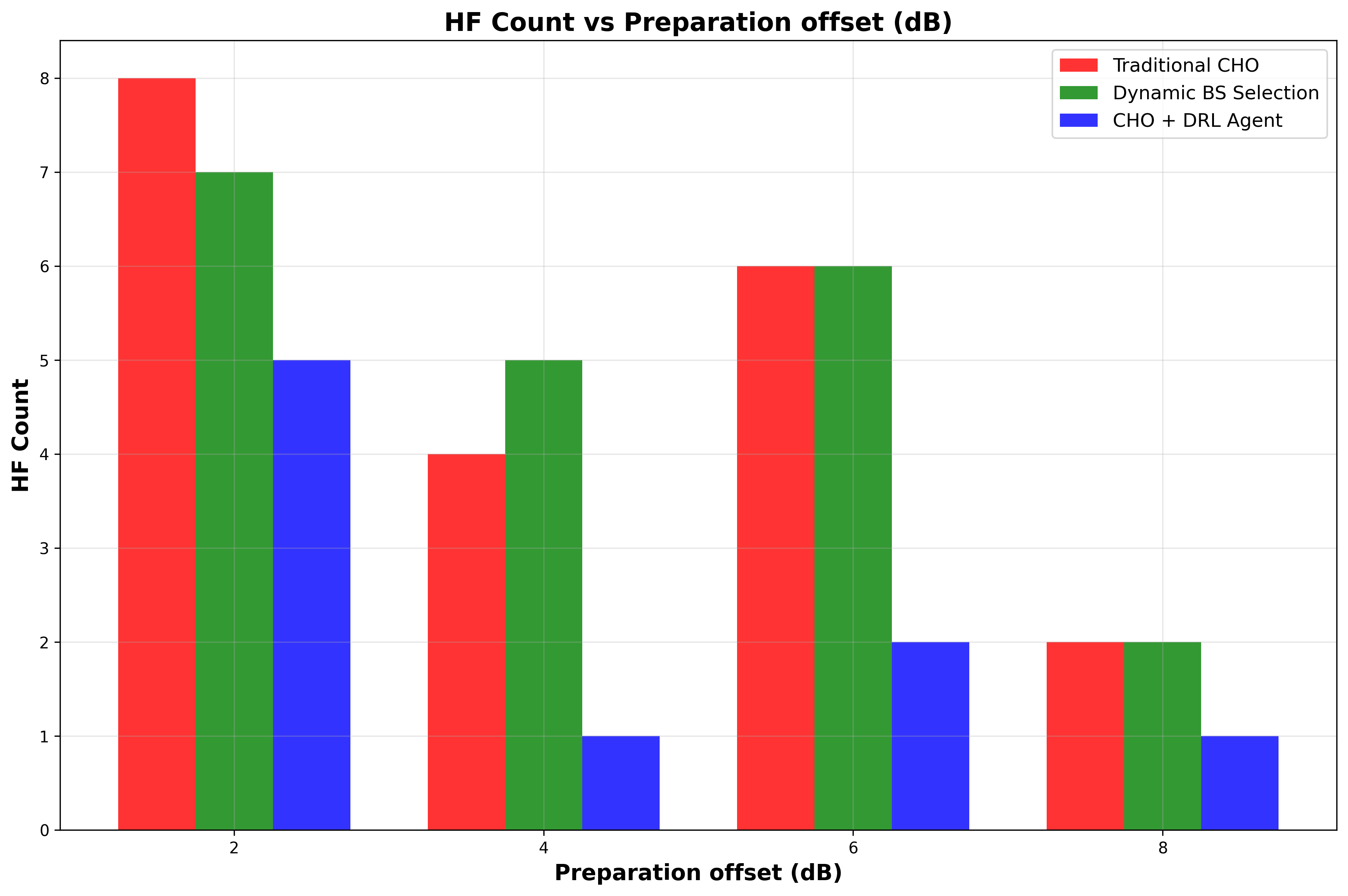}} \\
  \subfloat[HF Count vs. T310]{\includegraphics[width=5.5cm, height=5cm]{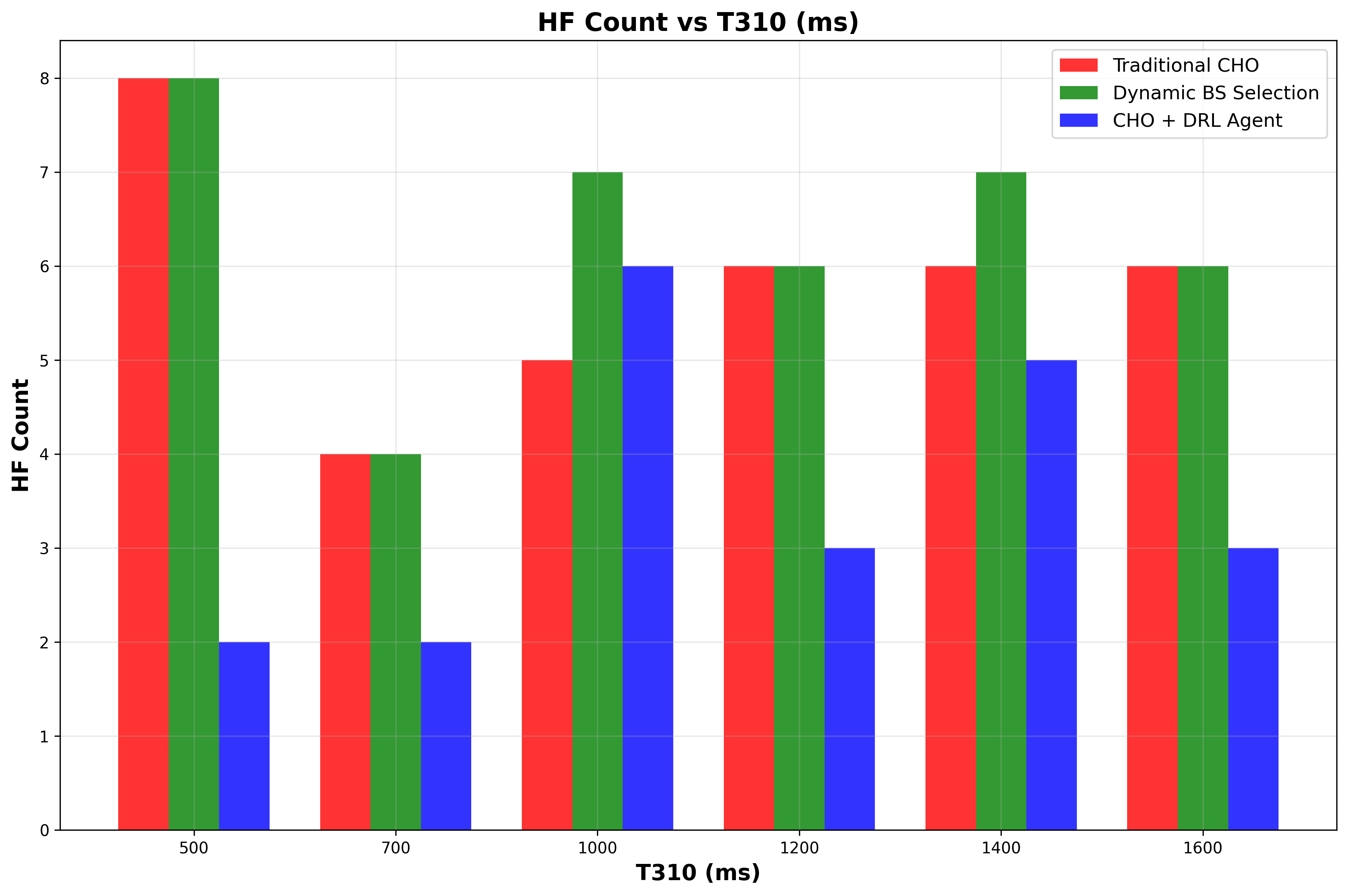}}
  \subfloat[HF Count vs. $T_{exec}$]{\includegraphics[width=5.5cm, height=5cm]{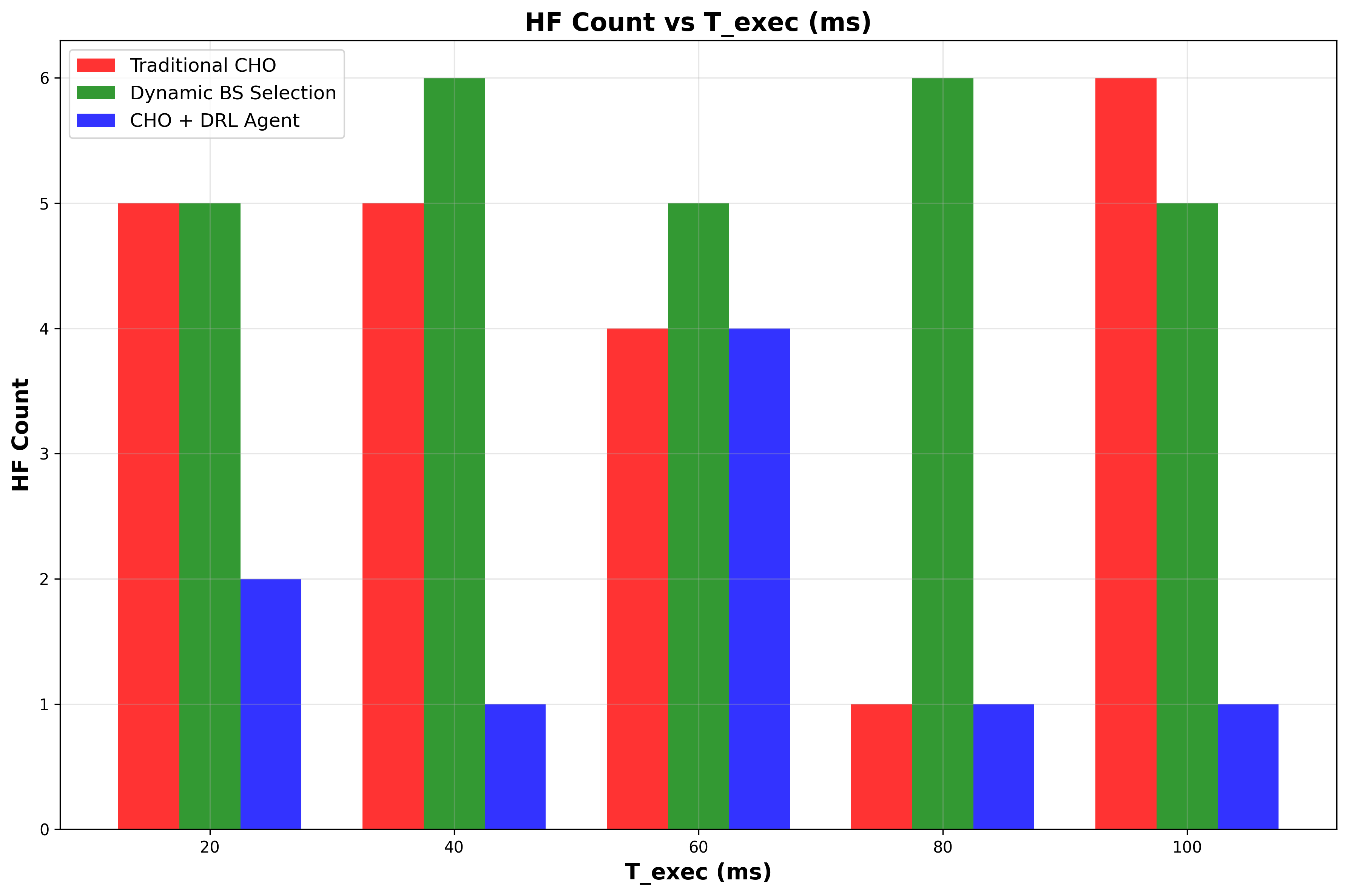}}
  \subfloat[HF Count vs. $T_{prep}$]{\includegraphics[width=5.5cm, height=5cm]{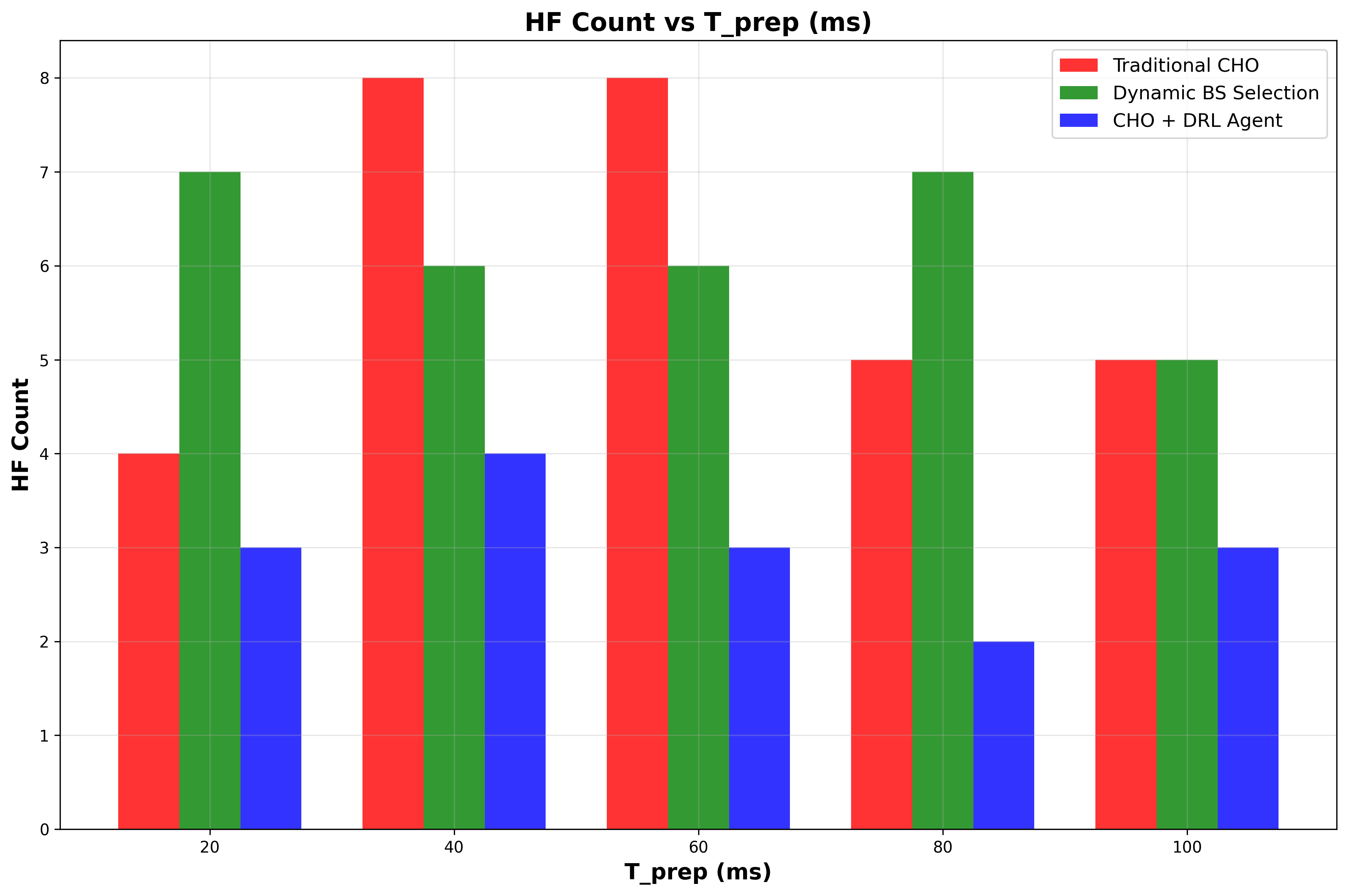}}
  \caption{Total HF counts under CHO and CHO+DRL across different parameter sweeps.}
  \label{fig:total_ho_count}
\end{figure*}

\begin{table}[ht]
  \caption{Coordinates and placement rationale of gNBs}
  \label{tab:enb_positions}
  \centering
  \begin{tabular}{c l l}
    \hline
    \textbf{ID} & \textbf{Location (x,y) [m]} & \textbf{Comment} \\
    \hline
     1  & (50, 150) & south of road, baseline coverage \\
     2  & (190, 330) & north of road, early handover trigger \\
     3  & (550, 180) & just south of road, weaker transition \\
     4  & (700, 290) & just north of road, coverage black spot \\
     5  & (880, 160) & south of road, artificial weak zone \\
     6  & (1040, 330) & north of road, standard spacing \\
     7  & (1190, 210) & slightly south, standard spacing \\
     8  & (1330, 260) & just north of road, weak transition area \\
     9  & (1630, 285) & north of road, intended handover zone \\
    10  & (1770, 155) & south of road, increased drop chance \\
    11  & (1940, 400) & far north, edge‑of‑coverage \\
    12  & (2230, 180) & south of road, offset from main corridor \\
    13  & (2385, 340) & north of road, just outside corridor \\
    14  & (2630, 115) & well south of road, past gap \\
    15  & (2830, 300) & north of road, high‑elevation weak link \\
    \hline
  \end{tabular}
\end{table}

\begin{table}[ht]
  \caption{Pairwise distances between consecutive gNBs}
  \label{tab:pairwise_distances}
  \centering
  \begin{tabular}{l c @{\hskip 1 cm} l c}
    \hline
    \textbf{Pair (IDs)} & \textbf{Distance [m]} & \textbf{Pair (IDs)} & \textbf{Distance [m]} \\
    \hline
    1–2   & 228.04  & 8–9   & 301.04 \\
    2–3   & 390.00  & 9–10  & 191.05 \\
    3–4   & 186.01  & 10–11 & 298.20 \\
    4–5   & 222.04  & 11–12 & 364.01 \\
    5–6   & 233.45  & 12–13 & 222.77 \\
    6–7   & 192.09  & 13–14 & 332.64 \\
    7–8   & 148.66  & 14–15 & 272.44 \\
    \hline
  \end{tabular}
\end{table}

\subsection{Simulation results}
In this section, we investigate the efficacy of the proposed DQN based power control mechanism in minimizing RLF and subsequent HF. We investigate the parametric impact of T310, N310, $O_{prep}$, $O_{exec}$, $T_{exec}$ and $T_{prep}$. Henceforth, the conventional CHO \cite{CHODeb} equipped with the proposed DQN based power control mechanism is referred to as CHO+DRL.

Figs.~\ref{fig:rlf_comparison}(a) and \ref{fig:total_ho_count}(a) show the impact of N310 on RLF and HF counts, respectively. The N310 threshold determines how many consecutive out‑of‑sync indications trigger the T310 timer for RLF detection. As N310 increases, RLF detection is deferred, limiting the agent’s opportunity to respond in time. At N310 = 5, our CHO+DRL agent achieves a 37.5\% reduction in RLF and 40\% reduction in HF compared to baseline, and outperforms the dynamic BS selection strategy~\cite{ericsson_dynamic_bs} by 58.3\% (RLF) and 62.5\% (HF) \textemdash demonstrating the agent’s ability to proactively boost power before timer expiry. While all methods suffer at higher thresholds due to delayed reaction, CHO+DRL maintains relative superiority.

Figs.~\ref{fig:rlf_comparison}(b) and \ref{fig:total_ho_count}(b) depict performance with respect to $O_{exec}$, the execution offset. The $O_{exec}$ indicates the required RSRP gap for the initiation of handover execution phase. Result shows that CHO when equipped with the proposed power control,  consistently outperforms the conventional CHO and the RL based dynamic BS selection approach towards minimizing RLF and HF counts. Maximum performance gain in terms of HF reduction is attained at $4$ dB (75\% drop in HF count); whereas performance gain in terms of RLF reductions are attained at $8$ dB (35\% drop in RLF count with 30\% drop in HF count). This is because RLF induced HF increases with increasing $O_{exec}$. Hence, the proposed power control mechanism become more effective with higher values of $O_{exec}$. The dynamic BS selection approach considers difference between the RSRP of the  serving  and the newly chosen gNB, therefore cannot predict an upcoming RLF. As a result our proposed algorithm outperforms the dynamic BS selection algorithm.  

Figs.~\ref{fig:rlf_comparison}(c) and \ref{fig:total_ho_count}(c) present results for varying preparation offset ($O_{prep}$). At $O_{prep} = 4$ dB, CHO+DRL reduces HF by 75\% vs. conventional CHO and 80\% vs. the RL based dynamic BS selection approach. RLF reductions peak at $O_{prep} = 2$ dB, with a 30\% gain over baseline and 25\% over the RL based dynamic BS selection approach. The DQN agent effectively suppresses premature handovers and counters signal loss during offset delays by learning environment-specific thresholds. This is because, lower $O_{prep}$ value triggers frequent handovers which are often unsuccessful and causes ping-pong effect. The DQN policy can prevent these premature handovers. As the $O_{prep}$ value increases, chance of RLF driven HF also decreases. In such cases, the DQN agent increases the transmitting power to avoid the RLF driven HF events. 

Figs.~\ref{fig:rlf_comparison}(d) and \ref{fig:total_ho_count}(d) show the effect of T310. The T310 timer governs the time duration before declaring an RLF. Result shows that the proposed power control mechanism can significantly reduce RLF and HF counts as compared to the considered state of the art approaches. The power control mechanism achieves an average RLF reductions of 50\% and HF reductions of 54.5\%, peaking at T310 = $1200$ ms. This is because, the agent increases power to prevent link failures during extended poor‐signal intervals.
  
Figs.~\ref{fig:rlf_comparison}(e) and \ref{fig:total_ho_count}(e) analyze the effect of T$_{exec}$. CHO+DRL reduces RLFs by 60\% vs. conventional CHO and RL based dynamic BS selection approach (peaking at $T_{exec} = 40$ ms). Result shows that the performance of CHO is consistently better when equipped with the power control agent. Figs.~\ref{fig:rlf_comparison}(f) and \ref{fig:total_ho_count}(f) analyze $T_{prep}$. Maximum RLF and HF reduction is observed at $T_{prep} = 60$ ms and $80$ ms, respectively. These results indicate that the agent effectively schedules power boosts during critical handover periods, thus maximizing link robustness.  

\section{Conclusion} \label{con}
In this work, a DQN based power control mechanism has been proposed which considers both handover and RLF parameters to minimize RLF driven HFs. The proposed approach has been compared with two state of the art approaches. Results show that the  conditional handover when equipped with the power control mechanism can significantly reduce RLFs and subsequent HFs. Our future research plan includes predicting the quantity of the power increase needed by the DRL agent in order to further minimize HF. Our future research scope includes the following:
\begin{itemize}
    \item Modifying the agent's action space to jointly adjust RLF parameters, handover parameters and transmit power. Allowing environment aware updates of the said parameters enables the agent to identify the optimal parameter values to minimize RLF driven HF under varying channel conditions.
    \item Redesigning the reward function and state space of the agent to capture the instantaneous and momentary attenuation in RSRP caused by dynamic obstacles. Consequently, an action to maintain current power for an adaptive duration will be added to avoid premature handovers initiated by dynamic obstacles.
    \item Augmenting the agent's state space with attributes that characterize different radio access technologies (RAT) such as millimeter wave (mmWave) and newly emerging THz communication. This enables an extension of the proposed algorithm for handovers in mmWave-THz heterogeneous networks.
    \item Extension of the proposed approach for reconfigurable intelligent surface (RIS) assisted networks by modifying the action space to jointly control transmit power from the gNB and RIS configuration to minimize RLF driven HF. 
\end{itemize}

\section*{Declaration}
A portion of this work has been submitted to the Shiv Nadar Institute of Eminence, Delhi NCR, India as an internal project report (OUR project number: OUR20240016).
\bibliographystyle{IEEEtran}
\bibliography{references}

\balance
\end{document}